\documentclass[11pt]{article}
\usepackage[margin=1in]{geometry}
\usepackage[T1]{fontenc}
\usepackage{lmodern}
\usepackage{microtype}

\usepackage{amsmath,amssymb,bm}
\usepackage{graphicx}
\usepackage{hyperref}
\usepackage{placeins}

\newcommand{\DKL}{D_{\mathrm{KL}}}

\newcommand{\keywords}[1]{%
  \begin{center}
  \small\textbf{Keywords:} #1
  \end{center}
  \vspace{0.5em}
}

\begin{document}

\hypersetup{hidelinks,
  pdftitle={Exchange-Symmetric Dissipation at the DNA--RNA Polymerase Interface},
  pdfauthor={Tatsuaki Tsuruyama},
  pdfsubject={Preprint}
}

\title{Exchange-Symmetric Dissipation at the DNA--RNA
Polymerase Interface}

\author{%
Tatsuaki Tsuruyama$^{1,2}$\\[0.4em]
\small $^1$Graduate School of Science, Tohoku University, Sendai, Japan\\
\small $^2$Department of Drug Discovery Medicine, Graduate School of Medicine, Kyoto University, Kyoto, Japan
}
\date{}

\maketitle


\begin{abstract}
In the DNA-RNA polymerase complex, RNA polymerase (RNAP) selectively incorporates ribonucleoside triphosphates according to the base information encoded in DNA, elongates the RNA strand, and progresses along the template. The state of the template DNA therefore conditions RNAP binding and chemical transitions. Conversely, from the viewpoint of DNA, the binding state and coordinate position of RNAP condition the local structural response of DNA. As RNAP advances, upstream DNA reanneals, downstream DNA opens, the RNA-DNA hybrid constrains the local geometry, and local bending and twisting of the DNA are updated.

We introduce an interface dissipation, $\Sigma_{\mathrm{int}}$, that provides an exchange-symmetric measure of path irreversibility at the DNA-RNAP interface, independent of whether DNA or RNAP is assigned the role of observer or observed. Irreversibility is quantified by the path-space Kullback-Leibler (KL) divergence between the forward path measure and its time-reversed counterpart, from which the dissipation of the joint system and those of the marginal systems are defined. The non-negativity of the interface dissipation derived in this work does not rely on RNAP-specific kinetic assumptions; rather, it holds for general path processes for which the relevant KL divergences and conditional path measures are well defined. RNAP transcription is treated as a biophysical application of this general theorem to the DNA-RNAP interface.
\end{abstract}

\section{Introduction}

\subsection{RNAP and information thermodynamics: relation to previous work}

RNA polymerase (RNAP) is an enzyme that moves along DNA by effective nucleotide-scale steps, incorporates the corresponding ribonucleotides, and elongates the RNA chain in the $5^\prime \to 3^\prime$ direction. Single-molecule and kinetic studies have resolved elongation, pausing, translocation, and backtracking over biologically relevant timescales \cite{Adelman2002,KimLarson2007,Turtola2018}.

\begin{figure}[htbp]
\centering
\includegraphics[width=0.95\linewidth]{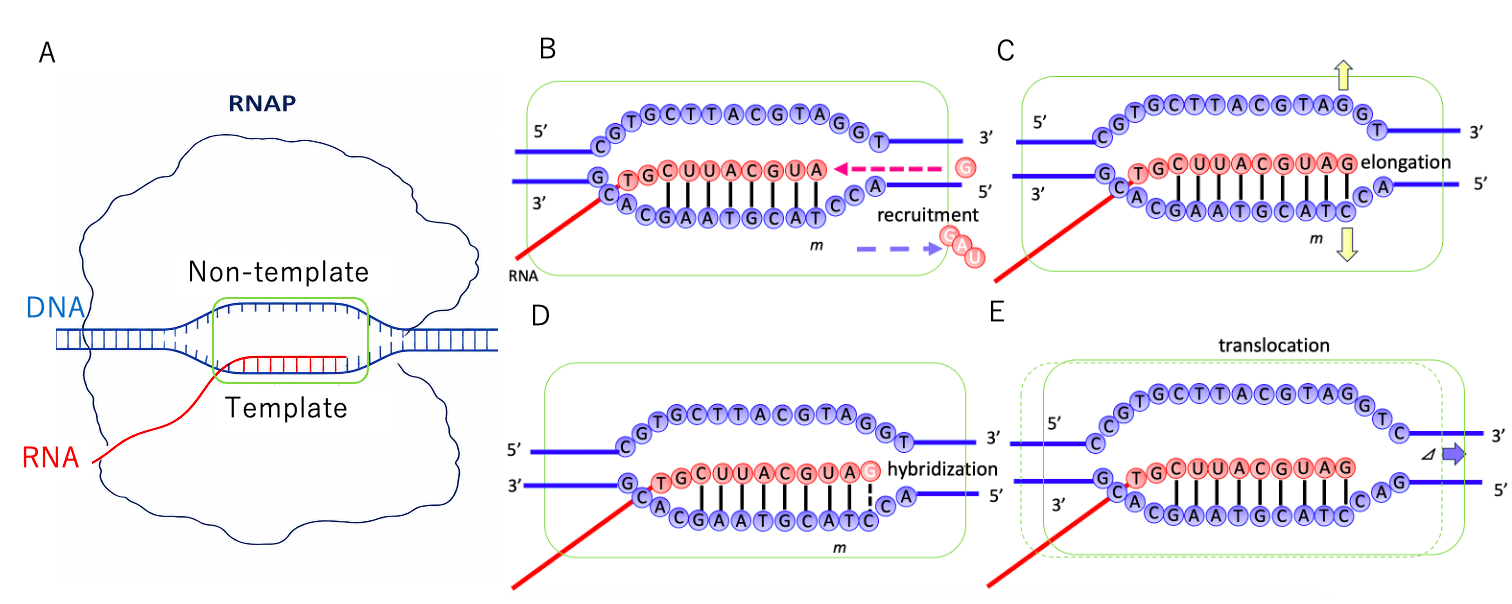}
\caption{
Conceptual diagram of the RNAP--DNA--RNA complex (A) and schematic representation of recruitment (B), elongation (C), hybridization, and translocation (D) during a one-nucleotide step. Blue lines and letters in circles represent the DNA strand and deoxyribonucleotides, respectively. Red lines and letters in circles represent the RNA strand and ribonucleotides, respectively.
}
\label{fig:rnap_model}
\end{figure}

RNAP transcription has also been discussed in an information-thermodynamic picture in which template information conditions polymerase dynamics \cite{Tsuruyama2023}. The present formulation uses an explicit product partition of the composite state rather than assigning the same transcription coordinate to both subsystems. In the CTMC benchmark below, the template register is assigned to the DNA-side variable \(X\), whereas the RNAP-side variable \(Y\) contains the internal binding and chemical state. This bookkeeping choice is made only to define non-overlapping projections of the same composite process; the exchange-symmetric construction itself is independent of which labels are assigned to the two factors. A schematic molecular model is shown in figure~\ref{fig:rnap_model}.

As RNAP advances, the transcription bubble moves forward. Upstream DNA reanneals, downstream DNA opens, and the position of the RNA--DNA hybrid shifts simultaneously. From the DNA side, RNAP translocation appears as a structural response involving opening of the double helix, local bending and twisting, and reannealing.

From the viewpoint of information thermodynamics, RNAP can be interpreted as a molecular information processor: it reads the template base and selectively binds and incorporates rNTPs in a base-dependent manner. In Brownian-ratchet descriptions of transcription, chemical selection and incorporation bias otherwise reversible translocational fluctuations toward productive elongation \cite{GuoSousa2006}. The Maxwell-demon analogy is therefore used here in the information-thermodynamic sense of state-dependent feedback, not as an external controller \cite{HorowitzEsposito2014,HartichBaratoSeifert2014}.

The bookkeeping of information and irreversibility depends on how the composite process is partitioned into two observed subsystems. In this paper, ``interface dissipation'' is an operational name for the exchange-symmetric combination of joint and marginal path-space irreversibilities defined in section~\ref{sec:theory}. It quantifies a symmetrized conditional time-reversal asymmetry of the chosen DNA--RNAP partition. As clarified below, it is not a pure interaction information and need not vanish when two irreversible subsystems are statistically independent.

This quantity is not, in general, identical to heat or to directly measured thermal dissipation. Only when local detailed balance holds, and when the logarithmic ratio of transition probabilities can be related to thermodynamic entropy production, can interface dissipation be interpreted as an interfacial component of physical entropy production or thermodynamic cost. Many existing information-thermodynamic quantities, such as learning rates and information flows, are defined with an orientation that depends on which part is taken as the observer and which as the observed. In contrast, the present work formulates the irreversibility associated with the DNA--RNAP interface in a form invariant under the exchange of these roles.

\subsection{Aim and main result of this work}

We first consider a general bipartite path process
\[
Z_t=(X_t,Y_t),
\]
and quantify irreversibility by path-space KL divergence. Here $X_t$ and $Y_t$ are path degrees of freedom obtained by partitioning an arbitrary composite system into two parts. In the application to RNAP, they correspond to the DNA-side path and the RNAP-side path, respectively. From the dissipation of the joint system, $\Sigma_{XY}$, and the marginal dissipations, $\Sigma_X$ and $\Sigma_Y$, we define an exchange-invariant interface dissipation below. Marginal dissipation refers to the irreversibility observed when only one part of the composite system is retained.

The central result of this paper is that the interface dissipation is invariant under the exchange
\[
X \leftrightarrow Y
\]
and is non-negative. This exchange-symmetric non-negativity relation is therefore not specific to an RNAP kinetic model; it applies to general bipartite path processes for which the relevant forward/reverse KL divergences and conditional path measures are well defined.

After establishing this general theorem, we interpret RNAP transcription as a concrete application. From the RNAP side, the DNA-side state $X$ acts as an input that conditions rNTP binding, unbinding, and chemical locking. Conversely, from the DNA side, the RNAP-side state $Y$ acts as an input that conditions local structural responses such as transcription-bubble motion, reannealing, opening, bending, and twisting. Thus the exchange $X \leftrightarrow Y$ is not an exchange of the molecules themselves, but an exchange of the partition used to describe the composite system.

A minimal continuous-time Markov chain (CTMC) is presented in appendix~\ref{app:ctmc_model} as a supplementary model that concretizes this general bipartite path process for a one-nucleotide RNAP step. Appendix~\ref{app:discrete_estimation} then connects the path-space quantities to discretely sampled trajectories. For the full sampled CTMC, the exact irreversibility rate is obtained independently from the sampled transition matrix \(P^{(\Delta t)}=\exp(Q\Delta t)\), whereas the projected DNA-side and RNAP-side processes are treated as hidden Markov processes. This provides an independent benchmark for assessing finite-context estimates. The Markov order \(r\) used in that appendix is therefore not a theoretical assumption of the non-negativity theorem, but an operational context length that controls a bias--variance trade-off when estimating projected path irreversibility from finite data.

\subsection{Relation to previous work}

The interface dissipation introduced here belongs to the standard framework of stochastic thermodynamics in the sense that it measures time-reversal asymmetry by path-space KL divergence. In bipartite systems, related decompositions have been formulated in terms of information flow, learning rate, transfer entropy, and Maxwell-demon-type inequalities \cite{HorowitzEsposito2014,HartichBaratoSeifert2014}. These quantities are intrinsically useful when one subsystem is identified as the information-acquiring or feedback-acting subsystem.

The present construction addresses a different question. Decompositions based on information flow or learning rate generally require an oriented assignment of which subsystem is the input and which is the response. In the RNAP example, DNA naturally appears as the input when one discusses rNTP selection and incorporation, whereas the RNAP state acts as an input when one describes transcription-bubble motion and the local structural response of DNA. We therefore define the interface quantity by treating the two marginal projections on an equal footing. This removes the directional ambiguity without replacing the oriented information-flow quantities.

The construction is also related to irreversibility under coarse-graining and hidden entropy production \cite{KawaguchiNakayama2013}. Hidden entropy production asks how much time-reversal asymmetry becomes invisible after eliminating degrees of freedom. The present \(\Sigma_{\mathrm{int}}\) addresses a different symmetrization: it averages the two non-negative gaps \(\Sigma_{XY}-\Sigma_X\) and \(\Sigma_{XY}-\Sigma_Y\). It therefore describes conditional irreversibility relative to the two projections, rather than a pure interaction contribution that would necessarily vanish in the absence of coupling.

Biochemical copying provides a second closely related body of work. Previous studies have analyzed kinetic versus energetic discrimination, proofreading, copying error, energetic cost, and nonequilibrium template--copy correlations \cite{SartoriPigolotti2013,SartoriPigolotti2015,OuldridgeGovernTenWolde2017,OuldridgeTenWolde2017,PoultonTenWoldeOuldridge2019}. These results make clear that copying accuracy, speed, and dissipation are distinct observables and need not vary monotonically with one another. The present work does not propose \(\Sigma_{\mathrm{int}}\) as a replacement for those quantities. Rather, it asks how to construct an exchange-symmetric conditional path-irreversibility measure for the chosen DNA--RNAP partition without privileging either observer--observed assignment.

To make the information-processing content of the RNAP application explicit, section~\ref{sec:ctmc_analysis} additionally introduces the mutual information between the template base and the incorporated rNTP. This copying-information quantity is distinct from \(\Sigma_{\mathrm{int}}\): the former measures template--copy correlation, whereas the latter measures exchange-symmetric time-reversal asymmetry. The non-negativity theorem itself does not require this additional information measure.

In summary, the theorem presented here is not derived from RNAP-specific kinetic assumptions. It is a general consequence of the chain rule for KL divergence in bipartite path processes. The RNAP model provides a biologically concrete application in which the exchange-symmetric irreversibility, copying fidelity, copying information, and physical current can be discussed as related but non-identical quantities.

\section{RNAP transcription and a one-nucleotide step: conceptual model and notation}

\subsection{Basic notation and the role of path processes}

The purpose of this section is to fix the minimal notation required for the path measures and interface dissipation introduced below. The main result of this paper is not intended to reproduce the full kinetic complexity of RNAP transcription. Rather, it is to make explicit the structure in which DNA-side and RNAP-side states are coupled through conditional path measures. Accordingly, a minimal CTMC representation in terms of specific rate constants and free energies is not required for the proof itself; it is summarized separately in appendix~\ref{app:ctmc_model}.

Let \(N_m\in\{A,C,G,T\}\) denote the identity of the template base at coordinate \(m\). We write
\[
\beta := (k_{\mathrm B}T)^{-1},
\]
where \(k_{\mathrm B}\) is the Boltzmann constant and \(T\) is the absolute temperature. In the following, the DNA-side path is denoted by \(X_{0:T}\), the RNAP-side path by \(Y_{0:T}\), and the joint path by
\[
(X_{0:T},Y_{0:T}).
\]
In the RNAP application, \(X\) and \(Y\) denote two non-overlapping projections of the composite DNA--RNAP state. In the explicit CTMC benchmark used below, \(X\) contains the template register (equivalently, the template-base identity for the periodic sequence), whereas \(Y\) contains the internal RNAP binding and chemical state. In a more detailed biophysical partition, additional structural degrees of freedom may be assigned to either factor provided that the two projections remain non-overlapping.

From the RNAP-side viewpoint, the DNA-side state \(X\) conditions the probability law of RNAP-side binding, unbinding, and chemical transitions. Conversely, the RNAP internal state can condition DNA-side structural responses such as transcription-bubble motion, reannealing, opening, bending, and twisting when those structural variables are retained in \(X\). The theoretical result below is derived from this conditional-path-measure structure.

\section{Theoretical formulation as a bipartite path process}
\label{sec:theory}

In this section, we describe the DNA--RNAP complex as a general bipartite path process and define interface dissipation based on path-space KL divergence. What is needed below is not the full set of CTMC transition rates, but the existence of the joint path measure \(P_{XY}\), its marginal path measures \(P_X\) and \(P_Y\), and the conditional path measures \(P_{Y|X}\) and \(P_{X|Y}\). Thus, the exchange-symmetric non-negativity relation does not depend on an RNAP-specific kinetic model. A minimal CTMC is given in appendix~\ref{app:ctmc_model} as a supplementary model that concretizes this abstract bipartite path process for a one-nucleotide RNAP step.

\subsection{Bipartite state space underlying the path measures}

The purpose of this subsection is to provide a bipartite state space that gives meaning to the path measures
\[
P_{XY},\qquad P_X,\qquad P_Y
\]
and to the conditional path measures
\[
P_{Y|X},\qquad P_{X|Y}.
\]
The state variables \(X_t\) and \(Y_t\) specify how the composite system is partitioned into two parts. They provide the basis for the projections used in the definition of the exchange-symmetric interface dissipation \(\Sigma_{\mathrm{int}}\).

For the explicit CTMC and estimator benchmark, we use a non-overlapping product partition. The DNA-side variable contains the template register,
\begin{equation}
X_t=m_t ,
\label{eq:ctmc_X_partition}
\end{equation}
which is equivalent to the template-base identity for the periodic ACGT sequence used below. The RNAP-side variable contains the internal binding/chemical state,
\begin{equation}
Y_t=(\chi_t,\rho_t),
\label{eq:ctmc_Y_partition}
\end{equation}
where \(\chi_t\in\{0,B,L\}\) and \(\rho_t\) denotes the rNTP identity when applicable. The pair \((X_t,Y_t)\) uniquely specifies the CTMC state. In a more detailed DNA-side description, \(X_t\) may additionally contain a coarse-grained structural variable \(q_t\) for transcription-bubble geometry, RNA--DNA hybrid constraints, or local bending and twisting. The theorem below does not depend on this particular partition; the explicit choice is stated only to make the projections used in the numerical benchmark unambiguous.

This bipartite structure naturally defines projections from the joint path
\[
(X_{0:T},Y_{0:T})
\]
onto the DNA-side path \(X_{0:T}\) and the RNAP-side path \(Y_{0:T}\). Therefore, \(\Sigma_{XY}\), \(\Sigma_X\), and \(\Sigma_Y\) below can be compared as joint and marginal path-measure quantities derived from the same composite process. The narrow bipartite Markov condition that each jump changes only one of the two degrees of freedom is convenient in explicit CTMC realizations, but it is not required for the non-negativity proof based on the KL chain rule. The proof applies to more general bipartite path processes as long as the relevant forward/reverse KL divergences and conditional path measures are well defined.

\subsection{RNAP transitions conditioned on DNA}

The joint path measure admits the two conditional factorizations
\begin{equation}
P_{XY}=P_XP_{Y|X}=P_YP_{X|Y}.
\label{eq:path_factorizations}
\end{equation}
Here \(P_{Y|X}\) is the probability law of the RNAP-side response path conditioned on the DNA-side history, while \(P_{X|Y}\) is the corresponding DNA-side conditional path law. In the RNAP application, \(P_{Y|X}\) expresses the base dependence of rNTP binding and incorporation. This is the operational sense in which RNAP ``reads'' the template: the transition law of the internal RNAP state depends on the template register. Correct nucleotide selection favors productive completion of the nucleotide-addition cycle, but the explicit CTMC below does not impose a direct asymmetry on the elementary thermal-translocation rates.

Conversely, a more detailed DNA-side model may describe transcription-bubble opening and closing or other structural responses conditioned on the RNAP internal state through \(P_{X|Y}\). Equation~\eqref{eq:path_factorizations} therefore provides two complementary conditional descriptions of the same joint process.

Thus, the DNA-side path \(X_{0:T}\), the RNAP-side path \(Y_{0:T}\), and the joint path \((X_{0:T},Y_{0:T})\) are defined. In the following, we prove non-negativity of interface dissipation using only this path-measure decomposition structure, without relying on the detailed form of a particular CTMC.

\subsection{Path-space KL divergence and interface dissipation}

As discussed above, in the RNAP application, RNAP-side binding and chemical transitions are conditioned on the DNA-side state \(X\). The formulation below, however, requires only that the joint and marginal path measures of a bipartite process be defined. We construct an exchange-symmetric combination of joint and marginal irreversibilities and then use the KL chain rule to express it as an average of two conditional path-space divergences. This construction is independent of which side, \(X\) or \(Y\), is labeled observer or observed.

Let \(P_{XY}\) be the forward path measure of the composite system \((X,Y)\) over a fixed time interval \([0,T]\), and let \(P_{XY}^{\mathrm R}\) be the corresponding time-reversed path measure. The path-space irreversibility of the joint system is defined by
\begin{equation}
\Sigma_{XY}
:=
\DKL\!\left(P_{XY}\,\middle\|\,P_{XY}^{\mathrm R}\right).
\label{eq:sigmaXY}
\end{equation}
Likewise, by projecting the joint path onto \(X\) and \(Y\), we obtain the marginal path measures \(P_X\) and \(P_Y\), together with their time reversals \(P_X^{\mathrm R}\) and \(P_Y^{\mathrm R}\). The marginal dissipations are
\begin{equation}
\Sigma_X
:=
\DKL\!\left(P_X\,\middle\|\,P_X^{\mathrm R}\right),
\qquad
\Sigma_Y
:=
\DKL\!\left(P_Y\,\middle\|\,P_Y^{\mathrm R}\right).
\label{eq:sigmaX_sigmaY}
\end{equation}
Since projection is an information-discarding operation, the data-processing inequality for KL divergence gives
\begin{equation}
\Sigma_{XY}\ge \Sigma_X,
\qquad
\Sigma_{XY}\ge \Sigma_Y.
\label{eq:data_processing}
\end{equation}

We define the interface dissipation by subtracting, with equal weight, the irreversibility visible in the two marginal descriptions from the irreversibility of the joint system:
\begin{equation}
\Sigma_{\mathrm{int}}
:=
\Sigma_{XY}
-\frac{1}{2}\left(\Sigma_X+\Sigma_Y\right).
\label{eq:sigma_int_def}
\end{equation}
This definition is manifestly invariant under the exchange \(X\leftrightarrow Y\). For long-time stationary discussions we use the rate notation
\begin{equation}
\dot\Sigma_A
:=
\lim_{T\to\infty}\frac{\Sigma_A(T)}{T},
\qquad
A\in\{X,Y,XY,\mathrm{int}\},
\label{eq:dissipation_rate_definition}
\end{equation}
whenever the limit exists.

The term ``interface'' should be interpreted operationally. The quantity in equation~\eqref{eq:sigma_int_def} is a symmetrized conditional irreversibility, not a pure interaction excess. In particular, if \(X\) and \(Y\) are statistically independent in both the forward and reversed ensembles, then \(\Sigma_{XY}=\Sigma_X+\Sigma_Y\) and
\begin{equation}
\Sigma_{\mathrm{int}}
=
\frac{1}{2}\left(\Sigma_X+\Sigma_Y\right).
\label{eq:independent_limit}
\end{equation}
Thus \(\Sigma_{\mathrm{int}}\) need not vanish for independent irreversible subsystems. In the DNA--RNAP application it is used as an exchange-symmetric descriptor of conditional path irreversibility for the chosen interface partition.

\paragraph{Theorem: exchange-symmetric non-negativity relation.}

Assume that the relevant forward path measures are absolutely continuous with respect to their time-reversed counterparts and that regular conditional path measures exist. Then \(\Sigma_{\mathrm{int}}\) is invariant under the exchange \(X\leftrightarrow Y\) and satisfies
\begin{equation}
\Sigma_{\mathrm{int}}\ge 0.
\label{eq:exchange_second_law}
\end{equation}

\paragraph{Proof.}

Non-negativity follows immediately from the data-processing inequalities in equation~\eqref{eq:data_processing}: each marginal irreversibility is no larger than the joint irreversibility, so their average is also no larger than \(\Sigma_{XY}\).

The KL chain rule gives a useful conditional interpretation. For joint path measures \(P_{XY}\) and \(Q_{XY}\),
\begin{equation}
\DKL(P_{XY}\|Q_{XY})
=
\DKL(P_X\|Q_X)
+
\mathbb E_{P_X}
\!\left[
\DKL(P_{Y|X}\|Q_{Y|X})
\right].
\label{eq:kl_chain_rule}
\end{equation}
Setting \(Q_{XY}=P_{XY}^{\mathrm R}\), and denoting by \(P_{Y|X}^{\mathrm R}\) the conditional measure induced by the reversed joint path measure \(P_{XY}^{\mathrm R}\), gives
\begin{equation}
\Sigma_{XY}-\Sigma_X
=
\mathbb E_{P_X}
\!\left[
\DKL\!\left(P_{Y|X}\,\middle\|\,P_{Y|X}^{\mathrm R}\right)
\right]
\ge0.
\label{eq:gap_X_nonnegative}
\end{equation}
Exchanging \(X\) and \(Y\) similarly yields
\begin{equation}
\Sigma_{XY}-\Sigma_Y
=
\mathbb E_{P_Y}
\!\left[
\DKL\!\left(P_{X|Y}\,\middle\|\,P_{X|Y}^{\mathrm R}\right)
\right]
\ge0.
\label{eq:gap_Y_nonnegative}
\end{equation}
Averaging these two identities gives
\begin{equation}
\Sigma_{\mathrm{int}}
=
\frac{1}{2}
\mathbb E_{P_X}
\!\left[
\DKL\!\left(P_{Y|X}\,\middle\|\,P_{Y|X}^{\mathrm R}\right)
\right]
+
\frac{1}{2}
\mathbb E_{P_Y}
\!\left[
\DKL\!\left(P_{X|Y}\,\middle\|\,P_{X|Y}^{\mathrm R}\right)
\right]
\ge0.
\label{eq:sigma_int_conditional_kl}
\end{equation}
The exchange symmetry follows directly from equation~\eqref{eq:sigma_int_def}. We refer to this non-negativity relation as the exchange-symmetric second-law form for the chosen bipartite partition.

\paragraph{Physical meaning.}
Equation~\eqref{eq:sigma_int_conditional_kl} shows that \(\Sigma_{\mathrm{int}}\) is the average of two conditional path-space irreversibilities: the RNAP-side forward/reverse asymmetry conditioned on the DNA-side path and the DNA-side asymmetry conditioned on the RNAP-side path. This interpretation is symmetric under exchange of the two descriptions. In the RNAP application, the first term contains the base dependence of binding and incorporation, while the second term represents the converse conditioning of the DNA-side description by the RNAP state when such structural degrees of freedom are retained.

Because of the independent-process limit in equation~\eqref{eq:independent_limit}, \(\Sigma_{\mathrm{int}}\) should not be interpreted as irreversibility uniquely generated by coupling. It is instead an exchange-symmetric conditional irreversibility associated with the chosen partition. Likewise, it is not generally identical to heat or chemical work. A thermodynamic entropy-production interpretation requires additional assumptions such as local detailed balance and a justified identification of path-probability ratios with environmental entropy production.

\subsection{Observer--observed swap as a symmetry of description}

The exchange \(X\leftrightarrow Y\) should be understood as an exchange of the partition used to describe the composite path. From the viewpoint of information thermodynamics, the assignment of observer and observed is not intrinsically fixed.

This point is particularly transparent from the DNA-side perspective. In the usual description, the template register is the input and RNAP binding and chemical transitions are the response. Conversely, a more detailed model can treat the RNAP internal state as conditioning transcription-bubble opening, reannealing, hybrid geometry, bending, or twisting on the DNA side. The physical transcription coordinate is an interface register and must be assigned to one factor when a non-overlapping product partition is constructed; in the explicit CTMC benchmark it is assigned to \(X\). Other consistent partitions are possible, but a given calculation must state the assignment unambiguously.

Many existing information-thermodynamic quantities are intrinsically oriented. For example, learning rates and information flows generally require one to decide in advance which subsystem is measuring and which subsystem is being measured, and they are not in general invariant under exchange. By contrast, \(\Sigma_{\mathrm{int}}\) treats the two marginal projections on an equal footing and provides a role-symmetric combination of conditional irreversibilities.

The structural point emphasized here is therefore not a further subdivision of RNAP kinetics, but an exchange-symmetric formulation of conditional irreversibility in a bipartite path process under observer--observed swap. The DNA--RNAP complex provides a biologically interpretable example of this construction. The discussion up to this point constitutes the theoretical core of the paper. The explicit CTMC model and the estimation procedure from discrete time series are placed in the appendices as operational supplements connecting the theoretical quantities to simulations and experimental data.

A useful consequence of this separation is that the role of memory in the projected processes can be tested independently of the theorem. In appendix~\ref{app:discrete_estimation}, the full sampled CTMC is first-order Markov and therefore has an exact \(r=1\) population description, whereas the projected \(X\)- and \(Y\)-processes retain hidden-state memory. Increasing \(r\) can recover more of this projected irreversibility at the population level, but it simultaneously enlarges the number of contexts that must be estimated from finite data. The numerical benchmark therefore interprets \(r\) through a bias--variance trade-off rather than as a hierarchy in which a numerically larger estimate is automatically better.

\subsection{Illustrative CTMC analysis: chemical driving, copying information, and kinetic robustness}
\label{sec:ctmc_analysis}

We next use the minimal CTMC specified in appendix~\ref{app:ctmc_model} to examine how chemical driving and base discrimination affect directly computable steady-state observables. The CTMC analysis in this subsection is intended as a kinetic illustration rather than as a quantitative fit to a particular RNAP species. We emphasize an important distinction from the path-space theorem: the generator gives the irreversibility rate of the full Markov process directly, whereas the exact marginal path-space irreversibilities required for \(\Sigma_{\mathrm{int}}\) involve projected processes that need not themselves be Markovian. We therefore plot the full CTMC irreversibility rate as a chemical-driving benchmark and do not identify it with \(\dot\Sigma_{\mathrm{int}}\).

The CTMC contains, for each template coordinate \(m\), an unbound state \(0_m\), rNTP-bound states \(B_{m,\rho}\), and post-incorporation locked states \(L_{m,\rho}\). The elementary cycle is
\begin{equation}
0_m \rightleftharpoons B_{m,\rho}
\rightleftharpoons L_{m,\rho}
\rightleftharpoons 0_{m+1},
\label{eq:ctmc_cycle_main}
\end{equation}
where the last pair of transitions represents thermal translocation. In the baseline calculation the forward and backward thermal translocation rates are taken to be symmetric,
\begin{equation}
k_+^{\mathrm{tr}}=k_-^{\mathrm{tr}}=10~{\rm s}^{-1}.
\label{eq:baseline_translocation_main}
\end{equation}
This choice defines a neutral thermal-translocation reference; it should not be interpreted as implying equal forward- and backward-step probabilities during processive transcription. Experimentally, RNAP predominantly progresses forward, whereas backtracking is generally less frequent and can become significant, reaching approximately \(10\)--\(20\%\) in particular sequence contexts \cite{Turtola2018}. In the present model, the symmetric values in equation~\eqref{eq:baseline_translocation_main} refer only to the elementary thermal translocation pair. Net directionality arises from the complete kinetic cycle in equation~\eqref{eq:ctmc_cycle_main}, in which base-dependent rNTP selection, polymerization, chemical locking, and translocation redistribute otherwise reversible fluctuations into net forward progression.

We define the base-discrimination parameter \(\Delta\Delta G_{\mathrm{bind}}\) as the effective free-energy penalty assigned to a non-cognate pathway relative to the cognate pathway. A non-cognate binding or incorporation channel is suppressed by
\begin{equation}
\exp\!\left(-\frac{\Delta\Delta G_{\mathrm{bind}}}{k_{\mathrm B}T}\right).
\label{eq:discrimination_factor_main}
\end{equation}
Thus, \(\Delta\Delta G_{\mathrm{bind}}=0\) corresponds to no discrimination, while larger values progressively suppress non-cognate channels.

The baseline parameter set is summarized in table~\ref{tab:ctmc_parameters}. These values define coarse-grained kinetic rate scales, not microscopic constants fitted to a particular RNAP species. Nevertheless, their order of magnitude is compatible with experimentally observed transcriptional timescales. Single-molecule measurements of \textit{E.\ coli} RNAP have reported mean elongation velocities of approximately \(12~\mathrm{nt\,s^{-1}}\), with pause-free velocities of a similar order \cite{Adelman2002}, and measurements of T7 RNAP likewise place processive elongation on the scale of several to several tens of nucleotides per second depending on nucleotide concentration and experimental conditions \cite{KimLarson2007}. Because one transition in the present coarse-grained CTMC does not correspond one-to-one to one experimentally observed nucleotide-addition event, these measurements should not be used to identify the individual rate constants directly. Rather, they show that the illustrative \(5\)--\(10~\mathrm{s^{-1}}\) rate scales are within a biologically relevant kinetic order of magnitude.

For this reason, we do not tune the rate constants to reproduce a particular measured elongation velocity or backtracking frequency. Instead, the robustness analysis below asks whether the qualitative conclusions survive changes in the kinetic rate scales. This provides a stronger test of the illustrative model than a single parameter calibration.

\begin{table}[htbp]
\caption{
Baseline parameters used in the illustrative CTMC. The kinetic rates are coarse-grained rate scales; the chemical-driving and discrimination variables are given in units of \(k_{\mathrm B}T\). The values are not a fit to a specific RNAP experiment.
}
\label{tab:ctmc_parameters}
\begin{center}
\begin{tabular}{lll}
\hline
Parameter & Meaning & Value \\
\hline
\(k_{\mathrm{on}}^0\) & rNTP binding rate scale & \(5~{\rm s}^{-1}\) \\
\(k_{\mathrm{off}}\) & rNTP unbinding rate & \(5~{\rm s}^{-1}\) \\
\(k_{\mathrm{pol}}^0\) & polymerization rate scale & \(5~{\rm s}^{-1}\) \\
\(k_+^{\mathrm{tr}}, k_-^{\mathrm{tr}}\) & symmetric thermal translocation rates & \(10~{\rm s}^{-1},10~{\rm s}^{-1}\) \\
\(\Delta\mu_{\mathrm{chem}}/k_{\mathrm B}T\) & chemical driving & \(0\)--\(8\) \\
\(\Delta\Delta G_{\mathrm{bind}}/k_{\mathrm B}T\) & base-discrimination strength & \(0\)--\(6\) \\
DNA sequence & template used for illustration & periodic ACGT \\
\hline
\end{tabular}
\end{center}
\end{table}

\begin{figure}[htbp]
\centering
\includegraphics[width=0.95\linewidth]{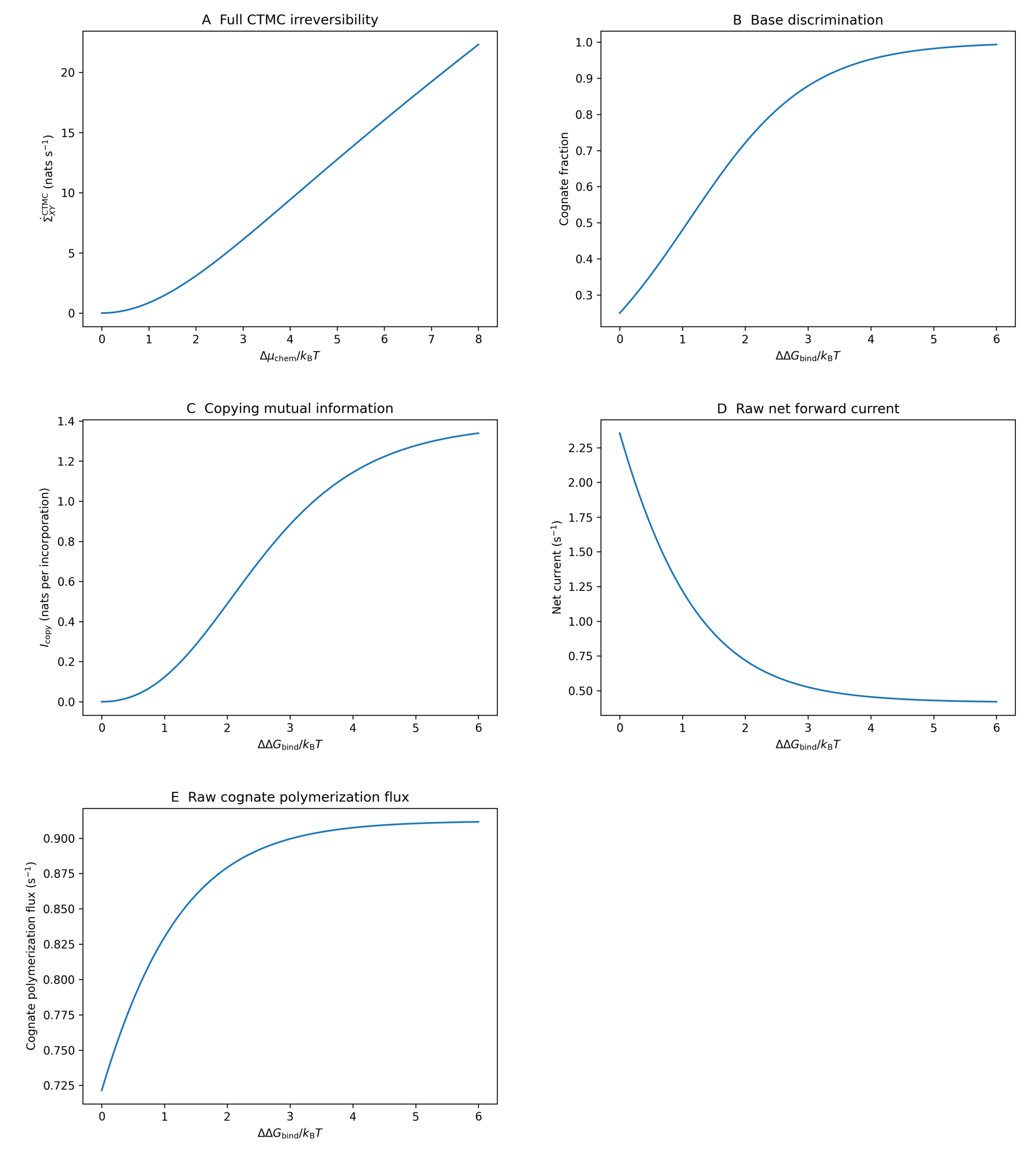}
\caption{
Illustrative CTMC analysis of chemical driving, copying information, and kinetic currents.
(A) Full CTMC irreversibility rate \(\dot\Sigma_{XY}^{\mathrm{CTMC}}\) versus chemical driving at \(\Delta\Delta G_{\mathrm{bind}}=0\).
(B) Cognate fraction versus base-discrimination strength at fixed \(\Delta\mu_{\mathrm{chem}}/k_{\mathrm B}T=4\).
(C) Copying mutual information \(I_{\mathrm{copy}}\) calculated directly from the stationary polymerization-flux distribution.
(D) Net forward translocation current.
(E) Cognate polymerization flux.
}
\label{fig:ctmc_analysis}
\end{figure}

\paragraph{Full-process irreversibility under chemical driving.}
For the stationary CTMC with stationary distribution \(\pi_i\) and transition rates \(w_{ij}\), we calculate the full-process irreversibility rate as
\begin{equation}
\dot\Sigma_{XY}^{\mathrm{CTMC}}
=
\sum_{i<j}
\left(\pi_i w_{ij}-\pi_j w_{ji}\right)
\ln
\frac{\pi_i w_{ij}}{\pi_j w_{ji}} .
\label{eq:ctmc_full_ep}
\end{equation}
At \(\Delta\Delta G_{\mathrm{bind}}=0\), this quantity is essentially zero at \(\Delta\mu_{\mathrm{chem}}=0\) and increases to \(3.09\), \(9.42\), \(16.04\), and \(22.32~\mathrm{nats\,s^{-1}}\) at \(\Delta\mu_{\mathrm{chem}}/k_{\mathrm B}T=2,4,6,\) and \(8\), respectively. Figure~\ref{fig:ctmc_analysis}A therefore provides a direct benchmark showing that increasing chemical driving increases the time-reversal asymmetry of the complete CTMC. This benchmark is deliberately denoted \(\dot\Sigma_{XY}^{\mathrm{CTMC}}\), rather than \(\dot\Sigma_{\mathrm{int}}\).

\paragraph{Copying information.}
To quantify explicitly how much information about the template is retained in the incorporated nucleotide, let \(J^{\mathrm{pol}}_{N\rho}\) denote the stationary forward polymerization flux for incorporation of rNTP \(\rho\) opposite template base \(N\). We define
\begin{equation}
p_{\mathrm{pol}}(N,\rho)
=
\frac{J^{\mathrm{pol}}_{N\rho}}
{\sum_{N',\rho'}J^{\mathrm{pol}}_{N'\rho'}}
\label{eq:polymerization_joint_distribution}
\end{equation}
and the copying mutual information
\begin{equation}
I_{\mathrm{copy}}
=
\sum_{N,\rho}
p_{\mathrm{pol}}(N,\rho)
\ln
\frac{p_{\mathrm{pol}}(N,\rho)}
{p_{\mathrm{pol}}(N)p_{\mathrm{pol}}(\rho)} .
\label{eq:copy_mutual_information}
\end{equation}
We also define the information flux
\begin{equation}
\dot I_{\mathrm{copy}}
=
J_{\mathrm{pol}} I_{\mathrm{copy}},
\qquad
J_{\mathrm{pol}}
=
\sum_{N,\rho}J^{\mathrm{pol}}_{N\rho}.
\label{eq:copy_information_flux}
\end{equation}
The quantity \(I_{\mathrm{copy}}\) measures template--copy correlation per incorporation event, whereas \(\dot I_{\mathrm{copy}}\) additionally weights this information by the incorporation throughput.

For the baseline rates at fixed \(\Delta\mu_{\mathrm{chem}}/k_{\mathrm B}T=4\), increasing \(\Delta\Delta G_{\mathrm{bind}}/k_{\mathrm B}T\) from \(0\) to \(6\) raises the cognate fraction from \(0.250\) to \(0.993\). Over the same range, \(I_{\mathrm{copy}}\) increases from zero to \(1.339\) nats per incorporation (\(1.93\) bits), approaching the four-symbol maximum \(\ln 4=1.386\) nats. At \(\Delta\Delta G_{\mathrm{bind}}/k_{\mathrm B}T=3\), the cognate fraction is \(0.879\) and \(I_{\mathrm{copy}}=0.884\) nats per incorporation. Thus, the base-discrimination parameter produces a quantitatively measurable increase in template--copy information, rather than merely a qualitative change in selectivity.

\paragraph{Speed, accuracy, and productive flux.}
The CTMC results clarify the apparently counterintuitive reduction of the net current. At the baseline parameter set, increasing \(\Delta\Delta G_{\mathrm{bind}}/k_{\mathrm B}T\) from \(0\) to \(6\) decreases the net forward current from \(2.355\) to \(0.420~\mathrm{s^{-1}}\), while the cognate fraction increases from \(0.250\) to \(0.993\). This occurs because weak discrimination leaves both cognate and non-cognate forward pathways open. Stronger discrimination removes the non-cognate contribution to total throughput; consequently, the total forward current can fall while the composition of that current becomes much more accurate. Mechanistically, correct rNTP selection and incorporation favor productive completion of the nucleotide-addition cycle and thereby convert otherwise reversible translocational fluctuations into directional progression. The model therefore attributes directionality to the coupled reaction cycle, not to an imposed asymmetry of the elementary thermal-translocation rates.

The cognate polymerization flux behaves differently from the total current. For the baseline rates it increases from \(0.721\) to \(0.912~\mathrm{s^{-1}}\) as the discrimination strength is increased from \(0\) to \(6\,k_{\mathrm B}T\). Thus, in the baseline kinetic regime, stronger discrimination reduces total throughput but redistributes the remaining productive traffic toward the correct channel. This distinction between total current, accurate productive flux, and copying information is the relevant speed--accuracy structure of the model.

\paragraph{Robustness to kinetic-rate changes.}
To test whether these trends are tied to the particular baseline rates, we performed a one-at-a-time sensitivity analysis. Each of \(k_{\mathrm{on}}^0\), \(k_{\mathrm{off}}\), and \(k_{\mathrm{pol}}^0\) was changed to \(0.5\) and \(2\) times its baseline value, while \(k_+^{\mathrm{tr}}\) and \(k_-^{\mathrm{tr}}\) were changed together by the same factors so that the symmetric thermal-translocation reference was preserved. For every perturbed parameter set, \(\Delta\mu_{\mathrm{chem}}/k_{\mathrm B}T\) was fixed at \(4\) and \(\Delta\Delta G_{\mathrm{bind}}/k_{\mathrm B}T\) was scanned from \(0\) to \(6\).

\begin{figure}[htbp]
\centering
\includegraphics[width=0.95\linewidth]{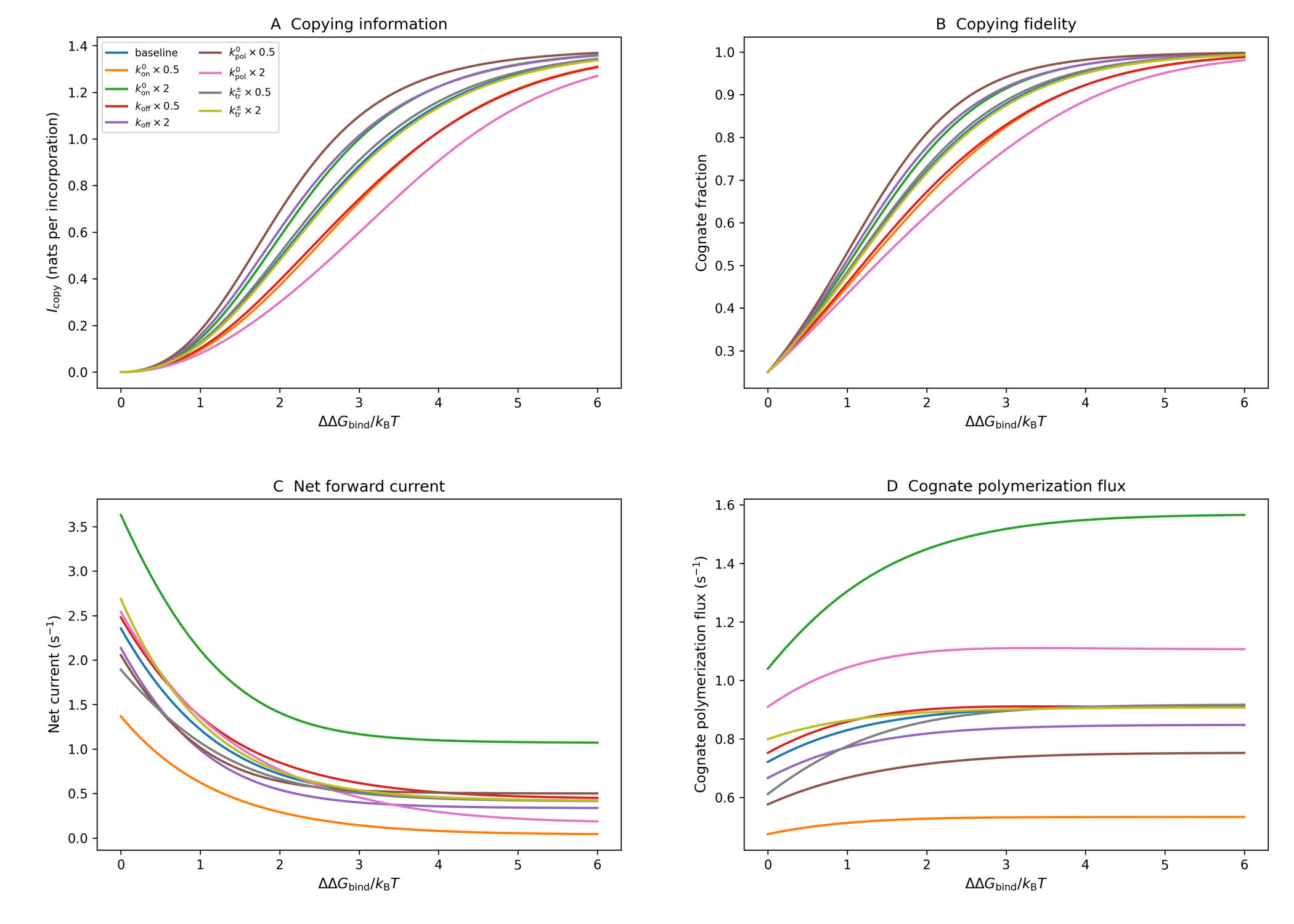}
\caption{
One-at-a-time kinetic robustness analysis at fixed
\(\Delta\mu_{\mathrm{chem}}/k_{\mathrm B}T=4\).
The binding, unbinding, and polymerization rate scales are independently varied by factors of \(0.5\) and \(2\), while the forward and backward thermal-translocation rates are varied together to preserve their baseline symmetry. The qualitative increase in \(I_{\mathrm{copy}}\) and cognate fraction and the decrease in net current are preserved throughout the tested range. The cognate flux generally increases and saturates, but can show a weak high-discrimination maximum for some rate choices.
}
\label{fig:parameter_robustness}
\end{figure}

Figure~\ref{fig:parameter_robustness} summarizes the sensitivity analysis. The most robust conclusions are the increase in copying information and fidelity and the decrease in total forward current. In all eight non-baseline perturbations, \(I_{\mathrm{copy}}\) and the cognate fraction increase monotonically with discrimination, whereas the net current decreases monotonically. At \(\Delta\Delta G_{\mathrm{bind}}/k_{\mathrm B}T=3\), the perturbed models give \(I_{\mathrm{copy}}=0.599\)--\(1.097\) nats, cognate fractions \(0.772\)--\(0.941\), and net currents \(0.143\)--\(1.167~\mathrm{s^{-1}}\). At \(6\,k_{\mathrm B}T\), \(I_{\mathrm{copy}}\) remains high, \(1.270\)--\(1.369\) nats, and the cognate fraction is \(0.981\)--\(0.998\).

By contrast, a strictly monotonic increase of the cognate polymerization flux is not universal. It is monotonic for six of the eight non-baseline perturbations, while for \(k_{\mathrm{off}}=0.5\,k_{\mathrm{off}}^{(0)}\) and \(k_{\mathrm{pol}}^0=2\,k_{\mathrm{pol}}^{0(0)}\) it reaches a shallow maximum near \(\Delta\Delta G_{\mathrm{bind}}/k_{\mathrm B}T\simeq3.3\) and then decreases by less than \(0.4\%\) by \(6\,k_{\mathrm B}T\). In both cases the flux at \(6\,k_{\mathrm B}T\) nevertheless remains above its value at zero discrimination. We therefore regard the monotonic gain in fidelity and copying information, together with the reduction of total throughput, as the robust result; the detailed high-discrimination behavior of the cognate flux depends weakly on the relative kinetic timescales.

The numerical analysis should therefore not be interpreted as a universal monotonic relation among information, speed, and dissipation. Instead, base discrimination robustly increases the information carried by the incorporated nucleotide and the probability of correct incorporation, while reducing the availability of error-producing forward channels. The detailed partition of the remaining flux depends on the kinetic timescales. This is consistent with the broader distinction between accuracy, speed, and energetic or path-space irreversibility in biochemical copying \cite{SartoriPigolotti2013,SartoriPigolotti2015,OuldridgeGovernTenWolde2017}.

\paragraph{Connection to path-space estimation.}
The CTMC generator used above also provides a controlled test of the operational estimator in appendix~\ref{app:discrete_estimation}. At the representative condition
\(\Delta\mu_{\mathrm{chem}}/k_{\mathrm B}T=4\),
\(\Delta\Delta G_{\mathrm{bind}}/k_{\mathrm B}T=3\), and
\(\Delta t=0.01~\mathrm{s}\), the continuous-time joint irreversibility is
\(3.3546~\mathrm{nats\,s^{-1}}\), while the exact irreversibility of the sampled joint chain is
\(3.3379~\mathrm{nats\,s^{-1}}\). The projected DNA-side and RNAP-side processes are non-Markovian after coarse-graining; an independent hidden-Markov path-likelihood calculation gives an interface reference of approximately
\(2.1544~\mathrm{nats\,s^{-1}}\).
At the same parameter point, the polymerization-event information flux from equation~\eqref{eq:copy_information_flux} is \(\dot I_{\mathrm{copy}}\simeq0.905~\mathrm{nats\,s^{-1}}\), whereas the sampled-path interface reference is approximately \(2.1544~\mathrm{nats\,s^{-1}}\). Their different values emphasize that copying information and conditional path irreversibility are distinct quantities; no universal equality between them is assumed. These independently known values allow the finite-context estimators to be judged by deviation from a reference rather than by their numerical ordering. The detailed convergence analysis, including \(r=1,2,3\) and finite-sample error bars, is given in figure~\ref{fig:finite_context_convergence} and appendix~\ref{app:discrete_estimation}.

\FloatBarrier

\section{Discussion}

The central theoretical result is the exchange-symmetric non-negativity of the quantity defined in equation~\eqref{eq:sigma_int_def}. The result follows from standard KL data processing and the chain rule and does not require an RNAP-specific kinetic model. Equation~\eqref{eq:sigma_int_conditional_kl} identifies the quantity as the average of two conditional path irreversibilities. As noted in equation~\eqref{eq:independent_limit}, it is not a pure coupling excess and can remain nonzero for independent irreversible subsystems; its role here is as an exchange-symmetric descriptor of the chosen DNA--RNAP partition.

The CTMC analysis separates this theoretical quantity from three other observables that are often conflated with it: the irreversibility of the full Markov process, copying accuracy, and copying information. The full CTMC rate \(\dot\Sigma_{XY}^{\mathrm{CTMC}}\) is directly obtainable from the generator and provides a benchmark for the effect of chemical driving. The copying mutual information \(I_{\mathrm{copy}}\), by contrast, measures template--copy correlation in polymerization events. Neither quantity is, by itself, identical to \(\Sigma_{\mathrm{int}}\). Exact evaluation of the marginal path-space terms entering \(\Sigma_{\mathrm{int}}\) requires the projected DNA-side and RNAP-side path measures; after projection these processes can have memory even when the full CTMC is Markovian.

The numerical benchmark in appendix~\ref{app:discrete_estimation} makes this distinction quantitative. The exact sampled joint process provides a known reference, while the projected path measures are evaluated independently with hidden-Markov likelihoods. For the representative condition used there, the exact sampled joint rate is \(3.3379~\mathrm{nats\,s^{-1}}\) and the projected-path interface reference is approximately \(2.1544~\mathrm{nats\,s^{-1}}\). The population finite-context estimates for the projected marginals increase with \(r\), because longer contexts recover additional hidden-state memory. Since \(\Sigma_{\mathrm{int}}\) subtracts these marginal contributions from the joint rate, the corresponding population interface estimates decrease toward the hidden-Markov reference. At finite sample size, however, the rapidly growing context space produces an opposing positive estimation bias and larger variance for \(r=2,3\). Thus the appropriate meaning of a ``better'' estimator is smaller deviation from an independently defined reference together with controlled sampling uncertainty, not a larger numerical value.

The explicit calculation of \(I_{\mathrm{copy}}\) strengthens the information-thermodynamic interpretation of the RNAP example. At zero discrimination the four rNTP channels are statistically symmetric and \(I_{\mathrm{copy}}=0\). Increasing \(\Delta\Delta G_{\mathrm{bind}}\) suppresses non-cognate channels, increases the cognate fraction, and drives \(I_{\mathrm{copy}}\) toward the four-symbol maximum \(\ln4\). Thus, the statement that RNAP ``reads'' the template is represented quantitatively by a growing template--copy correlation rather than only by a qualitative conditioning of transition rates.

The robustness analysis also clarifies the speed--accuracy interpretation. Across two-fold changes of the principal kinetic rate scales, increasing discrimination always raises the cognate fraction and \(I_{\mathrm{copy}}\) and always lowers the total net forward current in the tested model. The increase of the cognate polymerization flux seen for the baseline parameters is less universal: it can saturate and show a very shallow decline at strong discrimination for some rate ratios. Accordingly, the robust physical statement is not that stronger discrimination must increase every productive-current measure. Rather, discrimination closes error-producing channels, improves fidelity and copying information, and changes the total throughput; the detailed allocation of flux among productive pathways depends on the kinetic timescales.

The relation of \(\Sigma_{\mathrm{int}}\) to energetic cost still requires care. Its definition uses path-space time-reversal asymmetry. A thermodynamic entropy-production interpretation additionally requires assumptions such as local detailed balance and an appropriate identification of rate ratios with heat or chemical work. In coarse-grained descriptions with hidden states, the safer interpretation is operational: \(\Sigma_{\mathrm{int}}\) quantifies the symmetrized conditional irreversibility of the partitioned path ensemble.

The CTMC is therefore used here for three complementary purposes: to provide a concrete kinetic realization of the DNA--RNAP coupling, to demonstrate quantitatively how chemical driving, base discrimination, copying information, and physical currents respond to parameter changes, and to provide a controlled benchmark for path-space inference from finite trajectories. The theorem itself remains independent of these particular rate choices and of the finite-context estimator. Future extensions include explicit pausing and backtracking, misincorporation and proofreading networks, force and concentration dependence, and a systematic study of how the projected path-space rates \(\Sigma_X\), \(\Sigma_Y\), and \(\Sigma_{\mathrm{int}}\) depend on the observation partition and temporal resolution.

\appendix

\section{Model specification for the illustrative CTMC analysis}
\label{app:ctmc_model}

The exchange-symmetric non-negativity relation proved in the main text does not
depend on a specific kinetic model. It holds for general bipartite path
processes for which the relevant conditional path measures are defined.
In section~\ref{sec:ctmc_analysis}, however, we use an illustrative continuous-time Markov
jump process (CTMC) to provide a concrete kinetic realization of chemical
driving, base discrimination, copying information, and translocation
currents. This appendix specifies that model. The CTMC is not intended as
a fit to a particular experiment. Its generator directly determines the
irreversibility of the full Markov process. The marginal path-space
irreversibilities entering \(\Sigma_{\mathrm{int}}\), however, refer to
projected path measures and need not reduce to Markov-generator formulas
after coarse-graining; their operational estimation is discussed separately
in appendix~\ref{app:discrete_estimation}.

For the periodic template used in the numerical benchmark, the DNA-side
state \(X\) is represented by the template register \(m\), equivalently by
the corresponding template-base identity. The RNAP-side state \(Y\)
contains the rNTP binding state, incorporation state, and internal chemical
state, but not a duplicate copy of the register. Thus the pair \((X,Y)\)
forms the non-overlapping product partition stated in
equations~\eqref{eq:ctmc_X_partition} and \eqref{eq:ctmc_Y_partition}. In the calculation
involving base discrimination in section~\ref{sec:ctmc_analysis}, we distinguish the rNTP
identity
\[
\rho\in\{A,C,G,U\}.
\]
The RNAP-side state consists of an unbound state \(0\), rNTP-bound
states \(B_\rho\), and post-incorporation locked states \(L_\rho\).

Let \(\rho_c(N)\) denote the cognate rNTP for the template base \(N\).
The complementarity convention is
\begin{equation}
\rho_c(A)=U,\qquad
\rho_c(T)=A,\qquad
\rho_c(C)=G,\qquad
\rho_c(G)=C .
\label{eq:rc_complementarity}
\end{equation}

Cognate rNTP binding and incorporation proceed with the reference rates,
whereas non-cognate rNTP pathways are suppressed by the
base-discrimination strength \(\Delta\Delta G_{\mathrm{bind}}\).
We encode this suppression by the factor
\begin{equation}
\alpha_\rho(N)
=
\begin{cases}
1, & \rho=\rho_c(N),\\[4pt]
\exp\!\left[-\Delta\Delta G_{\mathrm{bind}}/(k_{\mathrm B}T)\right],
& \rho\ne \rho_c(N).
\end{cases}
\label{eq:alpha_discrimination}
\end{equation}
Thus, when \(\Delta\Delta G_{\mathrm{bind}}=0\), all rNTP pathways are
allowed with comparable weight. As \(\Delta\Delta G_{\mathrm{bind}}\)
increases, non-cognate rNTP binding and incorporation are increasingly
suppressed.

Using this notation, the binding and unbinding process is represented by
\begin{equation}
0
\mathrel{
\overset{k_{\mathrm{on}}^0\alpha_\rho(N)}
{\underset{k_{\mathrm{off}}}{\rightleftharpoons}}
}
B_\rho .
\label{eq:binding_process}
\end{equation}
The polymerization and reverse reaction are represented by
\begin{equation}
B_\rho
\mathrel{
\overset{k_{\mathrm{pol}}}
{\underset{k_{\mathrm{depol}}}{\rightleftharpoons}}
}
L_\rho .
\label{eq:polymerization_process}
\end{equation}

Chemical driving \(\Delta\mu_{\mathrm{chem}}\) is introduced through
the ratio between polymerization and the reverse reaction. With
\(k_{\mathrm{pol}}^0\) denoting the reference polymerization rate, the
illustrative calculation uses
\begin{equation}
k_{\mathrm{pol}}
=
k_{\mathrm{pol}}^0
\alpha_\rho(N)
\exp\!\left[
\frac{\Delta\mu_{\mathrm{chem}}}{2k_{\mathrm B}T}
\right],
\label{eq:kpol}
\end{equation}
and
\begin{equation}
k_{\mathrm{depol}}
=
k_{\mathrm{pol}}^0
\exp\!\left[
-\frac{\Delta\mu_{\mathrm{chem}}}{2k_{\mathrm B}T}
\right].
\label{eq:kdepol}
\end{equation}
Thus, at \(\Delta\mu_{\mathrm{chem}}=0\), polymerization and the reverse
reaction have no chemical directional bias, whereas increasing
\(\Delta\mu_{\mathrm{chem}}\) favors the post-incorporation locked state.
The factor \(\alpha_\rho(N)\) is also included in \(k_{\mathrm{pol}}\) as
an effective representation of the reduced probability of non-cognate
rNTP incorporation.

To isolate the Brownian-ratchet effect due to chemical locking, the
baseline calculation takes thermal translocation to be symmetric:
\begin{equation}
k_+^{\mathrm{tr}} = k_-^{\mathrm{tr}} .
\label{eq:symmetric_translocation}
\end{equation}
This assumption does not mean that the net motion of RNAP is
directionless. Rather, it treats forward and backward fluctuations in the
absence of chemical locking as a neutral reference process. The purpose
is to examine how rNTP incorporation and chemical driving selectively
stabilize forward states. In the baseline calculation of section~\ref{sec:ctmc_analysis}, we
set
\begin{equation}
k_+^{\mathrm{tr}} = k_-^{\mathrm{tr}} = 10~{\rm s}^{-1}.
\label{eq:baseline_translocation_rate}
\end{equation}

This CTMC defines the DNA-side path \(X_{0:T}\), the RNAP-side path
\(Y_{0:T}\), and the joint path \((X_{0:T},Y_{0:T})\). The full joint
irreversibility \(\Sigma_{XY}\) is obtained directly from the CTMC
generator. The projected quantities \(\Sigma_X\) and \(\Sigma_Y\), and
hence \(\Sigma_{\mathrm{int}}\), are path-space quantities of the marginal
processes. Because projection can introduce memory, their exact evaluation
generally requires projected-path likelihoods rather than a reduced
first-order generator. Appendix~\ref{app:discrete_estimation} implements
this distinction explicitly for the numerical benchmark.

The exchange-symmetric non-negativity relation
\begin{equation}
\Sigma_{\mathrm{int}}\ge 0
\label{eq:ctmc_nonnegative}
\end{equation}
does not depend on the particular number of states, rate expressions, or
parameter values chosen here. These choices are used only to provide a
concrete RNAP application and to reproduce the illustrative numerical
analysis in section~\ref{sec:ctmc_analysis}.

\section{Operational estimation from discretely sampled trajectories}
\label{app:discrete_estimation}

This appendix explains how the KL-type dissipation rates defined in the
main text can be estimated from discretely sampled RNAP trajectory data
obtained in experiments or simulations. In the main text, the DNA-side
path \(X_{0:T}\), the RNAP-side path \(Y_{0:T}\), and the joint path
\((X_{0:T},Y_{0:T})\) are treated as continuous-time stochastic
processes. The dissipations
\[
\Sigma_X,\qquad
\Sigma_Y,\qquad
\Sigma_{XY},\qquad
\Sigma_{\mathrm{int}}
\]
are defined by KL divergences between forward and time-reversed path
measures. By contrast, this appendix is not part of the proof of
non-negativity. It provides an operational connection between the
theoretical path-space quantities and finite observed time series.

In actual observations, the position of RNAP, its internal state, or the
state of the DNA--RNAP complex is rarely recorded as a complete
continuous-time path. More often, one obtains a finite time series sampled
at a fixed interval. To estimate the path-space dissipation rates from
data, one therefore samples the continuous-time path as a discrete
sequence and compares the likelihoods of the forward and time-reversed
sequences.

The sampling interval \(\Delta t\) used here is the temporal resolution of
observation. It is not the physical time required for RNAP to advance by
one nucleotide. A one-nucleotide step is a state transition in which the
transcription coordinate changes as
\[
m\to m+1 .
\]
The time at which such a transition occurs is random and is determined by
the continuous-time stochastic dynamics involving rNTP binding,
unbinding, incorporation, translocation, pausing, and backward motion.
Hence, during a single sampling interval \(\Delta t\), RNAP may remain at
the same register, may undergo only internal-state changes, or may
undergo one or more nucleotide translocations.

Sampling the observed continuous-time trajectory at a fixed interval
\(\Delta t\) gives samples at times
\begin{equation}
t_n=n\Delta t,\qquad n=0,1,\ldots,N .
\label{eq:sampling_times}
\end{equation}
In the notation of the main text, observation of the DNA side alone gives
\[
X_n:=X(t_n),
\]
observation of the RNAP side alone gives
\[
Y_n:=Y(t_n),
\]
and simultaneous observation of both sides gives
\[
(X_n,Y_n):=(X(t_n),Y(t_n)).
\]
To treat these cases in a unified notation, we write the target sequence
as \(S_n\):
\begin{equation}
S_n :=
\begin{cases}
X_n, & \text{when estimating }\Sigma_X,\\[3pt]
Y_n, & \text{when estimating }\Sigma_Y,\\[3pt]
(X_n,Y_n), & \text{when estimating }\Sigma_{XY}.
\end{cases}
\label{eq:Sn_definition}
\end{equation}
Thus, \(S_n\) is not a new physical variable. It is a notational
placeholder specifying whether the DNA-side marginal path, the RNAP-side
marginal path, or the joint DNA--RNAP path is used for estimating the
KL-type dissipation rate.

After discretization, one deals not with the full continuous-time path
measure itself, but with the probability of the finite observed sequence
\begin{equation}
S_{0:N}=(S_0,S_1,\ldots,S_N).
\label{eq:finite_sequence}
\end{equation}
Let \(P(S_{0:N})\) denote the probability law of this discrete sequence.
By the chain rule,
\begin{equation}
P(S_{0:N})
=
P(S_0)
\prod_{n=0}^{N-1}
p(S_{n+1}\mid S_{0:n}),
\label{eq:discrete_chain_rule}
\end{equation}
where \(p(\cdot\mid\cdot)\) is the one-step conditional probability and
\[
S_{0:n}=(S_0,\ldots,S_n)
\]
is the history up to index \(n\).

We now clarify why a Markov order \(r\) appears in this appendix. The
non-negativity theorem in the main text is based on the existence of
conditional path measures \(P_{Y|X}\) and \(P_{X|Y}\) at the
path-measure level. It does not assume a finite Markov order \(r\). In
contrast, when one estimates dissipation rates from experimental data,
one must estimate conditional probabilities from finite discrete time
series. Conditioning on the entire past history \(S_{0:n}\) is not
practical for finite samples. Therefore, one introduces an approximation
that uses only the most recent \(r\) observations.

Moreover, even if the underlying joint process \((X_t,Y_t)\) is
Markovian, the marginal sequences \(X_n\) or \(Y_n\) obtained by
projecting onto one side need not be Markovian. This is a consequence of
coarse-graining. Here, coarse-graining means that many internal states or
reaction stages are aggregated into a smaller number of observed states
for the purposes of measurement or modeling. Hidden states are states
that exist and affect RNAP dynamics but are not directly visible in the
observed sequence. Examples include transient rNTP binding, short-lived
RNAP conformational changes, microscopic opening and closing of the
transcription bubble, fluctuations of the RNA--DNA hybrid, and precursors
of short pauses or backtracking events. Eliminating such unobserved
degrees of freedom can induce apparent memory in the observed sequence.

We therefore approximate the full history-dependent conditional
probability by a finite-context approximation of length \(r\):
\begin{equation}
p(S_{n+1}\mid S_{0:n})
\approx
p^{(r)}(S_{n+1}\mid S_{n-r+1:n}).
\label{eq:markov_r_approx}
\end{equation}
Here, \(r\) is the Markov order, or context length, of the discretely
observed sequence. It should not be interpreted as a number of
nucleotides or as a physical correlation length. The case \(r=1\)
corresponds to a first-order Markov approximation using only the
immediately preceding state \(S_n\), whereas \(r\ge 2\) uses a longer
history block \(S_{n-r+1:n}\).

Next, define the time-reversed sequence. For the discrete sequence
\(S_{0:N}\), we set
\begin{equation}
\widetilde S_n := S_{N-n}.
\label{eq:reversed_sequence}
\end{equation}
A hat denotes an empirical quantity fitted from finite data. The forward conditional model
\(\hat p_{\mathrm F}^{(r)}(\cdot\mid\cdot)\) is fitted from forward
training trajectories, whereas
\(\hat p_{\mathrm R}^{(r)}(\cdot\mid\cdot)\) is fitted from the
time-reversed training trajectories. Importantly, a forward/reverse
likelihood ratio compares the probability of the \emph{same observed
transition and context} under the two models. With
\[
C_n^{(r)}:=S_{n-r+1:n},
\]
the held-out finite-context estimator used below is therefore
\begin{equation}
\widehat{\dot\Sigma}^{(r)}[S]
=
\frac{1}{N_{\mathrm{test}}\Delta t}
\sum_{(C_n^{(r)},S_{n+1})\in{\cal D}_{\mathrm{test}}}
\log
\frac{
\hat p_{\mathrm F}^{(r)}
\!\left(S_{n+1}\mid C_n^{(r)}\right)
}{
\hat p_{\mathrm R}^{(r)}
\!\left(S_{n+1}\mid C_n^{(r)}\right)
}.
\label{eq:kl_rate_estimator}
\end{equation}
This form is the empirical conditional-KL rate between the forward and
time-reversed sequence laws. It differs from comparing a forward event
with a different event at the corresponding index of the reversed
sequence.

The long-time rate notation \(\dot\Sigma_A\) was defined in
equation~\eqref{eq:dissipation_rate_definition}. In this appendix, a hat
indicates a finite-data estimate and the superscript \((r)\) denotes the
finite context length.

Applying equation~\eqref{eq:kl_rate_estimator} to \(S=X\), \(S=Y\),
and \(S=(X,Y)\) yields
\[
\widehat{\dot\Sigma}_X^{(r)},\qquad
\widehat{\dot\Sigma}_Y^{(r)},\qquad
\widehat{\dot\Sigma}_{XY}^{(r)}.
\]
The interface estimator is
\begin{equation}
\widehat{\dot\Sigma}_{\mathrm{int}}^{(r)}
=
\widehat{\dot\Sigma}_{XY}^{(r)}
-
\frac{1}{2}
\left(
\widehat{\dot\Sigma}_X^{(r)}
+
\widehat{\dot\Sigma}_Y^{(r)}
\right).
\label{eq:sigma_int_estimator}
\end{equation}

\begin{figure}[htbp]
\centering
\includegraphics[width=\linewidth]{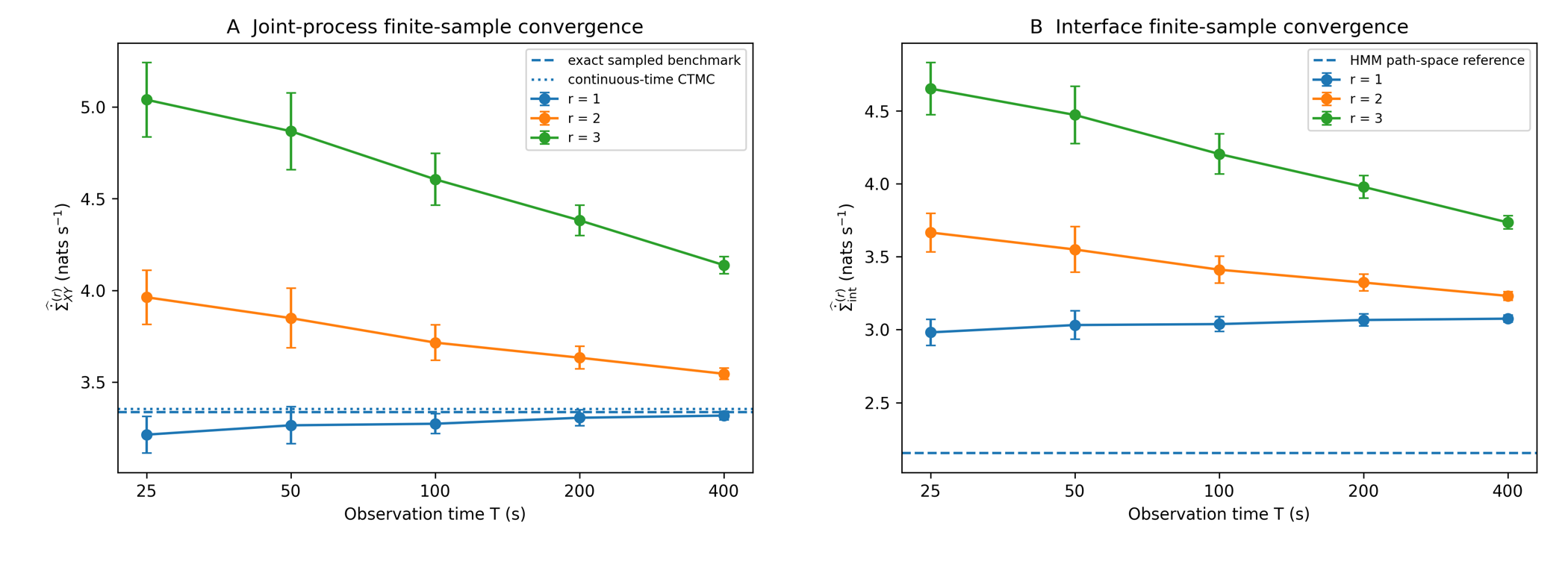}
\caption{
Finite-context benchmark at
\(\Delta\mu_{\mathrm{chem}}/k_{\mathrm B}T=4\),
\(\Delta\Delta G_{\mathrm{bind}}/k_{\mathrm B}T=3\), and
\(\Delta t=0.01~{\rm s}\).
Each point is the mean over \(20\) independent ensembles of \(100\)
trajectories; \(50\) trajectories are used to fit the forward and reverse
conditional models and \(50\) are held out for likelihood evaluation.
Error bars denote one standard deviation.
(A) Joint-process estimator. The dashed reference is the exact sampled
joint rate, \(3.3379~{\rm nats\,s^{-1}}\), and the dotted reference is the
continuous-time CTMC rate, \(3.3546~{\rm nats\,s^{-1}}\).
Because the full sampled process is first-order Markov, \(r=1\) is already
the correct population model. The positive finite-sample bias and variance
for \(r=2,3\) decrease as \(T\) increases.
(B) Interface estimator. The dashed line is the independent hidden-Markov
path-likelihood reference,
\(\dot\Sigma_{\mathrm{int}}^{\mathrm{HMM}}\simeq
2.1544~{\rm nats\,s^{-1}}\).
For \(r=1,2,3\) in this benchmark, increasing \(r\) reduces temporal truncation of the marginal processes at the
population level, but the larger context space produces stronger
finite-sample bias and variance. The two effects therefore act in opposite
directions at the observation times shown.
}
\label{fig:finite_context_convergence}
\end{figure}

For the numerical convergence test, the \(100\) trajectories in each
ensemble are divided into \(50\) training and \(50\) held-out trajectories.
The conditional probabilities are regularized by the Jeffreys
pseudocount,
\begin{equation}
\hat p(s\mid c)
=
\frac{N(c,s)+1/2}{N(c)+|\mathcal S|/2},
\label{eq:jeffreys_context}
\end{equation}
where \(N(c,s)\) is the number of occurrences of context \(c\) followed
by state \(s\) in the training data. This regularization avoids undefined
logarithms for unobserved finite-sample transitions. Twenty independent
trajectory ensembles are generated, and the error bars in
figure~\ref{fig:finite_context_convergence} show one standard deviation
across these ensembles.

\paragraph{Exact joint benchmark.}
Because the full process is a known CTMC, its continuous-time
irreversibility rate is available without a finite-context approximation:
\begin{equation}
\dot\Sigma_{XY}^{\mathrm{CTMC}}
=
\sum_{i<j}
\left(\pi_i w_{ij}-\pi_j w_{ji}\right)
\ln
\frac{\pi_i w_{ij}}{\pi_j w_{ji}}.
\label{eq:exact_ctmc_kl_rate}
\end{equation}
The discrete estimator is evaluated at a finite sampling interval, so the
most direct benchmark is the exact sampled Markov chain
\begin{equation}
P^{(\Delta t)}=\exp(Q\Delta t),
\label{eq:sampled_transition_matrix}
\end{equation}
whose stationary path-space irreversibility rate is
\begin{equation}
\dot\Sigma_{XY}^{(\Delta t)}
=
\frac{1}{\Delta t}
\sum_{i,j}
\pi_i P_{ij}^{(\Delta t)}
\ln
\frac{\pi_i P_{ij}^{(\Delta t)}}
{\pi_j P_{ji}^{(\Delta t)}}.
\label{eq:exact_sampled_kl_rate}
\end{equation}
For the representative benchmark condition
\[
\Delta\mu_{\mathrm{chem}}/k_{\mathrm B}T=4,\qquad
\Delta\Delta G_{\mathrm{bind}}/k_{\mathrm B}T=3,\qquad
\Delta t=0.01~{\rm s},
\]
equations~\eqref{eq:exact_ctmc_kl_rate} and
\eqref{eq:exact_sampled_kl_rate} give
\[
\dot\Sigma_{XY}^{\mathrm{CTMC}}
=3.3546~{\rm nats\,s^{-1}},
\qquad
\dot\Sigma_{XY}^{(\Delta t)}
=3.3379~{\rm nats\,s^{-1}}.
\]
The small difference is the irreversibility hidden by finite-time
sampling. Because the complete sampled CTMC is first-order Markov, the
population \(r=1\) model is already exact for the joint process; increasing
\(r\) adds parameters but no additional population-level information.

\paragraph{Projected-path reference and the meaning of ``better''.}
For the estimator benchmark we use the product partition
\(X=m\), or equivalently the periodic template base, and
\(Y=(\chi,\rho)\), where
\(\chi\in\{0,B,L\}\) is the RNAP internal state and \(\rho\) labels the rNTP
when applicable. The pair \((X,Y)\) uniquely specifies the \(36\)-state
CTMC state. After projecting onto either \(X\) or \(Y\), however, the
observed sequence is a hidden Markov process.

An independent reference for the projected path irreversibilities is
obtained from the exact sampled transition matrix
\(P^{(\Delta t)}\) and its stationary time reversal
\begin{equation}
P^{R}_{ij}
=
\frac{\pi_jP^{(\Delta t)}_{ji}}{\pi_i}.
\label{eq:reverse_sampled_transition}
\end{equation}
Forward and reverse hidden-Markov likelihoods are evaluated on a
\(2\times10^6\)-step stationary trajectory. This gives
\[
\dot\Sigma_X^{\mathrm{HMM}}\simeq0.8220,\qquad
\dot\Sigma_Y^{\mathrm{HMM}}\simeq1.5450
\quad {\rm nats\,s^{-1}},
\]
and hence
\begin{equation}
\dot\Sigma_{\mathrm{int}}^{\mathrm{HMM}}
=
\dot\Sigma_{XY}^{(\Delta t)}
-\frac{1}{2}
\left(
\dot\Sigma_X^{\mathrm{HMM}}
+\dot\Sigma_Y^{\mathrm{HMM}}
\right)
\simeq2.1544~{\rm nats\,s^{-1}}.
\label{eq:hmm_interface_reference}
\end{equation}
Ten additional \(3\times10^5\)-step trajectories give a standard error
of approximately \(0.007~{\rm nats\,s^{-1}}\) for this projected
interface reference.

The population finite-context rates make the two sources of error
explicit. For the projected marginals, increasing \(r\) from 1 to 3 in this benchmark gives a
tighter finite-context description of the hidden-memory irreversibility:
\[
\begin{array}{c|ccc}
r & 1 & 2 & 3\\ \hline
\dot\Sigma_X^{(r)} & 0.0492 & 0.0849 & 0.1403\\
\dot\Sigma_Y^{(r)} & 0.4373 & 0.5419 & 0.6436
\end{array}
\quad {\rm nats\,s^{-1}}.
\]
Thus, for this benchmark, retaining more history reveals additional
marginal irreversibility at the population level. However, the interface quantity is a
difference between the joint and marginal rates. The corresponding
population finite-context interface values are
\begin{equation}
\dot\Sigma_{\mathrm{int}}^{(1)}=3.0947,\qquad
\dot\Sigma_{\mathrm{int}}^{(2)}=3.0245,\qquad
\dot\Sigma_{\mathrm{int}}^{(3)}=2.9460
\quad {\rm nats\,s^{-1}},
\label{eq:interface_population_context}
\end{equation}
which decrease toward the hidden-Markov reference as more marginal memory
is retained. Consequently, a numerically larger
\(\widehat{\dot\Sigma}_{\mathrm{int}}^{(r)}\) is not a tighter lower
bound and is not, by itself, a ``better'' estimate.

The joint panel directly demonstrates finite-sample overestimation for the
higher-order plug-in models. At \(T=25~{\rm s}\), the mean
joint estimates for \(r=1,2,3\) are \(3.215\), \(3.963\), and
\(5.039~{\rm nats\,s^{-1}}\), respectively, compared with the exact sampled
value \(3.338~{\rm nats\,s^{-1}}\). At \(T=400~{\rm s}\), they are
\(3.318\pm0.023\), \(3.546\pm0.031\), and
\(4.138\pm0.047~{\rm nats\,s^{-1}}\), where the quoted uncertainties are
standard deviations across the \(20\) ensembles. Thus the excess of the
higher-order estimates decreases with increasing data length, while their
sampling variability also decreases.

For the interface estimator, finite-context truncation and finite-sample
error must be distinguished. In the infinite-data finite-context limit,
increasing \(r=1\to3\) lowers the interface estimate from \(3.095\) to
\(2.946~{\rm nats\,s^{-1}}\), toward the HMM reference \(2.154\).
At finite data length, however, the larger context spaces for \(r=2,3\)
produce a positive estimation bias that can dominate this population-level
improvement. This explains why the finite-sample curves can be ordered in
the opposite direction. In this appendix, ``better'' therefore means
smaller deviation from an independently defined reference at a stated
sampling resolution, together with smaller sampling uncertainty; it does
not mean a larger numerical value.

The important point is that neither the exchange-symmetric non-negativity relation nor
the CTMC analysis in the main text depends on this finite-context
estimation procedure. The Markov order \(r\) is not an assumption
required for the non-negativity of \(\Sigma_{\mathrm{int}}\). It is an
operational parameter used when estimating KL-type dissipation rates from
discretely sampled experimental time series. The theory in the main text
is based on path measures and the KL chain rule, whereas
appendix~\ref{app:discrete_estimation} provides a supplementary procedure
for connecting those theoretical quantities to experimental or simulation
data.

\FloatBarrier

\clearpage

\clearpage

\end{document}